\documentclass[american,aps,pra,reprint, superscriptaddress]{revtex4-1}
\usepackage{color}
\usepackage{times}
\usepackage{subfig}
\usepackage{bm}
\usepackage{graphicx}
\usepackage{graphics}
\DeclareGraphicsExtensions{.pdf,.png,.jpg}
\usepackage{amsbsy}
\usepackage{amsmath}
\usepackage{amsfonts}
\usepackage{amsthm}
\usepackage{float}
\usepackage{amssymb}
\usepackage{textcomp}
\usepackage{mathtools}
\usepackage{hyperref}
\usepackage[font=small,labelfont=bf,
justification= centerlast,
format=hang]{caption}

\usepackage[subnum]{cases}
\usepackage{enumitem}
\usepackage[font=small,labelfont=bf,
justification= centerlast,
format=hang]{caption}
\DeclareMathOperator\supp{sup}
\DeclareMathOperator{\Tr}{Tr}
\definecolor{purple(html/css)}{rgb}{0.5, 0.0, 0.5}
\newcommand{\ket}[1]{| #1 \rangle}

\newcommand{\ketbra}[2]{| #1 \rangle \langle #2 |}

\begin{document}

\title{A geometric way to find the measures of uncertainty from statistical divergences for discrete and finite probability distributions}

\author{Gautam Sharma}
\email{gautam.oct@gmail.com}
\affiliation{Optics and Quantum Information Group, Institute of Mathematical Sciences, HBNI, CIT Campus, Taramani, Chennai 600113, India}

\author{Sk Sazim}
\email{sk.sazimsq49@gmail.com}
\affiliation{RCQI, Institute of Physics, Slovak Academy of Sciences, Bratislava, 845 11, Slovakia}

\begin{abstract}
Exploiting the geometric nature of statistical divergences, we devise a way to define associated induced uncertainty measures for discrete and finite probability distributions. We also report new uncertainty measures and discuss their properties. Further, we apply a similar technique to measure the uncertainty in the preparation of a quantum state.
\end{abstract}

\maketitle
\section{To begin with}
Modern science has been going through a transformation with the inclusion of tools from information science. The information theoretic approach has led to describe physical phenomenon in a more operational way. A common brigde is the information measure of an event -- the entropy. 
After the seminal work by Shannon \cite{https://doi.org/10.1002/j.1538-7305.1948.tb01338.x}, a plethora of measures of entropy have been discovered and studied. Most of these constructions of entropies were built under the assumption of certain axioms \cite{Fad56,DIDERRICH1975149,e10030261}. One of the operational application of the entropic quantities have been to quantify the uncertainty in a measurement outcome.

Distance between two probability distributions captures the difference in information content between them. These distances are sometime called divergences because of their non-symmetric nature. It is well understood that for each divergence there exists an information measure. For example, consider Shannon entropy and Kullback-Liebler divergence 
\cite{e10030261}. One can obtain the Shannon entropy of a finite probability distribution $\mathcal{P}=\{p_i\}_1^n$, in terms of the Kullback-Liebler divergence \cite{kullback1951information} of $\mathcal{P}$ from the uniform distribution $\mathcal{P}_U=(1/n,...,1/n)$ \cite{6832827}, i.e.
\begin{align*}
	S(\mathcal{P})=\log n- D_{KL}(\mathcal{P}, \mathcal{P}_U),
\end{align*}
where $S(\mathcal{P})=-\sum_ip_i\log p_i$ is the Shannon entropy, $D_{KL}(\mathcal{P},\mathcal{P}_U)$ is the Kullback-Liebler divergence of $\mathcal{P}$ from $\mathcal{P}_U$, and the base of the logarithm is taken $2$ for the whole paper. 
However, such a derivation does not suggest why these quantities are a valid measures of information (uncertainty). 
It has been shown that all the non-negative Schur concave functions which take zero value for a maximally certain probability distributions are valid measures of uncertainty \cite{PhysRevLett.111.230401}. It is not always clear how to obtain an information measure for a given divergence measure.

In this work, we show using geometric approach how to obtain an uncertainty measure from statistical divergence measures. Our approach is universal and intuitive in the sense that we define the uncertainty measure of a probability distribution $\mathcal{P}$ from a maximally certain probability distribution. 

 The rest of the paper is organised as follows. In sec.\ref{sec2}, we define uncertainty as the distance from a maximally certain distribution. Then, we give examples of obtaining several known and new uncertainty measures in sec.\ref{sec3}. In sec.\ref{sec4}, we discuss the properties of new uncertainty measures that we found. We use a similar technique to quantify the uncertainty in the preparation of a quantum state in sec.\ref{sec5} and finally we conclude in sec.\ref{sec6}.

\section{Uncertainty as the statistical distance from the most certain distributions}\label{sec2}
Given a discrete probability distribution $\mathcal{P}=\{p_i\}_{i=1}^n$, we want to give an intuitive meaning to its uncertainty, with $\sum_ip_i=1$. A probability distribution has zero uncertainty iff one of the $p_i=1$. There can be $n$ number of such probability distributions, we label them as $\mathcal{P}_C$. On the other hand, a probability distribution is said to be maximally uncertain iff each $p_i=\frac{1}{n}$, labelled as $\mathcal{P}_U$. It is obvious that, there can only be a unique  $\mathcal{P}_U$. 

Intuitively, we know that the uncertainty in a probability distribution $\mathcal{P}$ will be large if $\mathcal{P}$ is very far from $\mathcal{P}_C$ or very close to $\mathcal{P}_U$. Thus, we can quantify uncertainty using a statistical divergence measure $D(\mathcal{P}||\mathcal{P}_C)$, from a maximally certain distribution $\mathcal{P}_C$. But there is an issue that there are $n$ possible distributions $\mathcal{P}_C$, so for a distribution $\mathcal{P}$ there will be $n$ possible distances $D(P||\mathcal{P}_C)$, where as we are looking for a unique value. To resolve this, we note that the maximally uncertain distribution $\mathcal{P}_U$, is unique and it is equidistant from all $\mathcal{P}_C$. Using this fact we employ the following expression to define uncertainty.
\begin{align}\label{disuncer}
U^{\uparrow}(\mathcal{P}) = D(\mathcal{P}_C||\mathcal{P}_U)- D(\mathcal{P}||\mathcal{P}_U).
	\end{align}

The above expression of $U^{\uparrow}(\mathcal{P})$ is uniquely defined and captures the distance of $\mathcal{P}$ from the set of $\mathcal{P}_C$ uniquely. Using any statiscal divergence measure $D(\mathcal{P}||\mathcal{Q})$ between two probability distributions $\mathcal{P}$ and $\mathcal{Q}$, one can get different measures of uncertainty. 
\begin{figure}[htb]
	\centering
	\includegraphics[scale=0.6]{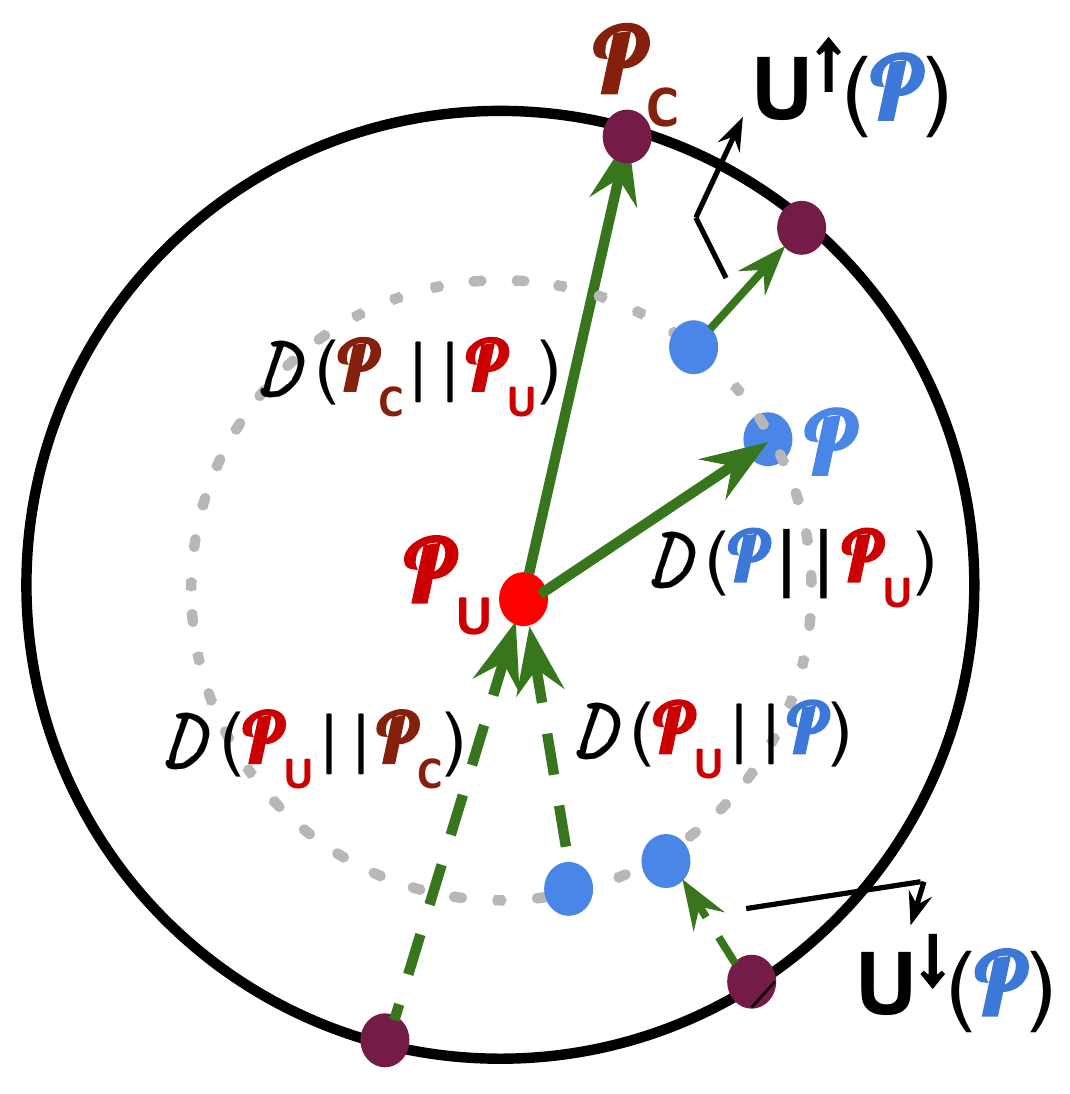}
	\caption{A schematic describing how the uncertainty measures are geometrically defined. All the points lying on a unique circle are equally uncertain. The divergences represented by vectors pointing away from the center give the uncertainty measure $U^{\uparrow}(\mathcal{P})$ where as the divergences represented by vectors pointing towards the center give $U^{\downarrow}(\mathcal{P})$ as the measure of uncertainty.}
	\label{fig:a}
\end{figure}

Further, a divergence measure need not be symmetric, i.e., $D(\mathcal{P}||\mathcal{Q})$ might not be equal to $D(\mathcal{Q}||\mathcal{P})$. Therefore, we can have another definition of uncertainty as 

\begin{align}\label{disuncer2}
	U^{\downarrow}(\mathcal{P}) = D(\mathcal{P}_U||\mathcal{P}_C)- D(\mathcal{P}_U||\mathcal{P}).
\end{align}

We explain the two measures from a schematic diagram in Fig.(\ref{fig:a}). The asymmetry of the divergence measures is depicted via vectors in opposite directions, one directed away from the center and other directed towards the center.

To ensure that the uncertainty measures $U^{\uparrow}(\mathcal{P})$ and $U^{\downarrow}(\mathcal{P})$ are valid uncertainty measures, they should be non-negative Schur-concave functions, see \cite{PhysRevLett.111.230401}. In Eq.(\ref{disuncer}) and Eq.(\ref{disuncer2}) the first terms $D(\mathcal{P}_C||\mathcal{P}_U)$ and $D(\mathcal{P}_U||\mathcal{P}_C)$ respectively, are constants. Therefore, the second terms $D(\mathcal{P},\mathcal{P}_U)$ and $D(\mathcal{P}_U,\mathcal{P})$ must be a Schur-convex function(as the negative of Schur-convex function is a Schur-concave function), to ensure that the quantities in Eq.(\ref{disuncer}) and Eq.(\ref{disuncer2}) are a valid uncertainty measures.

\section{Constructing uncertainty measures using f-divergences}\label{sec3}
In this section, we show how different statistical divergence measures can lead to various uncertainty measures. A category of divergence measures for which the second terms in Eq.(\ref{disuncer}) and Eq.(\ref{disuncer2}) are Schur-convex, are the \textit{f}-divergence measures \cite{10028997448,e10030261,10.2307/2984279}. This can be seen very easily as, for all the \textit{f}-divergences  $D(\mathcal{P}||\mathcal{P}_U)=\sum_{i=1}^{n}\frac{1}{n}f(n p_i)$, where $f:[0,\infty)\rightarrow \mathbb{R}$ is a convex function, which gives a Schur-convex function. Here we have used the fact that the linear sum of convex function leads to a Schur-convex function \cite{peajcariaac1992convex}. Using the same property we also have,  $D(\mathcal{P}_U||\mathcal{P})=\sum_{i=1}^{n}p_if(\frac{1}{n p_i})$, a Schur-convex function. 

Next, we give a few examples of constructing the uncertainty measures by substituting a well known f-divergences measures in  Eq.(\ref{disuncer}) and Eq.(\ref{disuncer2}). We will show, how both equations can lead to different measures whenever the given f-divergence is asymmetric.

\subsection{Renyi Divergence}
For two discrete probability distributions $\mathcal{P}=\{p_i\}_{i=1}^n$ and $\mathcal{Q}=\{q_i\}_{i=1}^n$, the Renyi Divergence is defined as \cite{renyi1961measures} 
\begin{align}\label{renyi}
D_{\alpha}(\mathcal{P}||\mathcal{Q})=\frac{1}{\alpha-1}\log \left(\sum_{i=1}^{n}\frac{p_i^{\alpha}}{q_i^{\alpha-1}}\right).
\end{align}

where $\alpha \in (0,1)\cup (1,\infty)$. Renyi divergence can also be defined for $\alpha=0,1$ and $\infty$ by taking a limit. 
\subsubsection{Measures from $U^{\uparrow}(\mathcal{P})$}
First we substitute the Renyi divergence in Eq.(\ref{disuncer}), which gives the well known Renyi Entropy\cite{1055890,aczel1975measures} as the uncertainty measure. 
\begin{align}\label{renyient}
U^{\uparrow}_{\alpha}(\mathcal{P})&= D_{\alpha}(\mathcal{P}_C||\mathcal{P}_U)- D_{\alpha}(\mathcal{P}||\mathcal{P}_U) \nonumber \\
&=\frac{1}{\alpha-1}\log n^{\alpha-1}-\frac{1}{\alpha-1}\log \left(n^{\alpha-1}\sum_i^np_i^{\alpha}\right)\nonumber \\
&=\frac{1}{1-\alpha}\log \sum_i^n p_i^{\alpha}.
\end{align}

In the following we mention the form of Renyi divergence and corresponding uncertainty measure for a few special values of $\alpha$.\\

\begin{itemize}
	\item For $\alpha=0$,  
     	  \\ \hspace*{1.3cm}  $D_{0}(\mathcal{P}||\mathcal{Q})=-\log Q$,
	     \\ \hspace*{1.3cm} $U^{\uparrow}_0(\mathcal{P})=\log Q.$

where $\mathcal{Q}$ is the cardinality of the probability space for which $p_i$ is non-zero. Thus, we get the Hartley/Max Entropy measure of uncertainty for $\alpha=0$.
	\item $\alpha=\frac{1}{2},   
\\ \hspace*{1.3cm} 	D_{\frac{1}{2}}(\mathcal{P}||\mathcal{Q})=-2\log \sum_{i=1}^{n}\sqrt{p_iq_i}
\\ \hspace*{1.3cm} 	 U^{\uparrow}_{\frac{1}{2}}(\mathcal{P})=2\log(\sum_{i=1}^{n}\sqrt{p_i})$.

In this case, the Renyi divergence becomes the negative log of ``Bhattacharya Coefficient", which is also a measure of overlap of probability distributions \cite{10.2307/25047882}.	
	\item $\alpha=1,  
	 \\ \hspace*{0.8cm}  D_{1}(\mathcal{P}||\mathcal{Q})=D_{KL}(\mathcal{P}||\mathcal{Q})=\sum_{i=1}^{n}p_i\log \frac{p_i}{q_i}
	 \\ \hspace*{1.3cm}  U^{\uparrow}_{1}(\mathcal{P})=-\sum_i^n p_i\log p_i$.
	 
In this case, Renyi divergence takes the form of ``Kullback-Leibler divergence" donoted as $D_{KL}(\mathcal{P}||\mathcal{Q})$, which gives the well known Shannon entropic measure of uncertainty. There also exists a symmetric form of Kullback-Leibler divergence, known as the Jensen-Shannon divergence, which we discuss in the next subsection. 
	\item $\alpha=\infty, 
	\\  \hspace*{1.3cm} D_{\infty}(\mathcal{P}||\mathcal{Q})=\log \frac{p_i}{q_i}_{max}
	\\ \hspace*{1.3cm}   U^{\uparrow}_{\infty}(\mathcal{P})=-\log {p_i}_{max}$.
	
	This is known as the min-Entropy, as it is the smallest in the family of Renyi entropies.
\end{itemize}

\subsubsection{Measures from $U^{\downarrow}(\mathcal{P})$}

Next, we substitute Renyi divergence in Eq.(\ref{disuncer2}), which gives the following measure of uncertainty.

\begin{align}
U^{\downarrow}_{\alpha}(\mathcal{P})&= D_{\alpha}(\mathcal{P}_U||\mathcal{P}_C)- D_{\alpha}(\mathcal{P}_U||\mathcal{P}) \nonumber \\
&= \frac{1}{\alpha-1}\log\left(\frac{1}{n^{\alpha}}\right)-\frac{1}{\alpha-1}\log\left(\sum_{i=1}^{n}\frac{1}{n^{\alpha}p_i^{\alpha-1}}\right) \nonumber \\
&=\frac{1}{1-\alpha}\log\left(\sum_{i=1}^{n}p_i^{1-\alpha}\right)\nonumber.
\end{align}

The above measure is well defined only for $\alpha \in (0,1)$. We can redefine $1-\alpha=\gamma$, so that 
\begin{align}
	U^{\downarrow}_\gamma(\mathcal{P})=\frac{1}{\gamma}\log\left(\sum_{i=1}^{n}p_i^{\gamma}\right).
\end{align}
It can be easily seen on comparing the above measure with the Renyi entropy that $U^{\downarrow}_{\alpha}(\mathcal{P})=\frac{\alpha-1}{\alpha}U^{\uparrow}_{\alpha}(\mathcal{P})$, i.e. the above measure is the rescaled Renyi Entropy measure for $\alpha\in(0,1)$.

\subsection{Jensen-Shannon Divergence}
The Jensen-Shannon divergence between two probability distributions, $D_{JS}(\mathcal{P}||\mathcal{Q})$ has the following form \cite{61115,MENENDEZ1997307}
\begin{align*}
	D_{JS}(\mathcal{P}||\mathcal{Q})&=D_{KL}(\mathcal{P}||\frac{\mathcal{P}+\mathcal{Q}}{2})+D_{KL}(\mathcal{Q}||\frac{\mathcal{P}+\mathcal{Q}}{2}). 
\end{align*}

As the Jensen-Shannon divergence is symmetric, it will give the same uncertainty measure via both Eq.(\ref{disuncer}) and Eq.(\ref{disuncer2}). On substituting $D_{JS}(\mathcal{P}||\mathcal{Q})$ in Eq.(\ref{disuncer}), we get the following measure of uncertainty which we denote as $U_{JS}(\mathcal{P})$
\begin{align}\label{jenshan}
	U_{JS}(\mathcal{P})=&\log\Big(\frac{4n^2}{(n+1)^{\frac{1+n}{n}}}\Big)-\sum_{i=1}^{n}p_i\log(p_i)\nonumber \\ &+\frac{2}{n}\sum_{i=1}^n\frac{np_i+1}{2}\log\Big(\frac{np_i+1}{2}\Big).
\end{align}
This entropy looks different than the usual Shannon entropy because of the last term. In next Section, we will discuss a few of its properties in details. Note also that one can consider more general Shannon-Jenson divergence by replacing KL-divergence with one parameter f-divergence (see \cite{e10030261}), which may induce a new information measure.

\subsection{Tsallis Divergence}
Again, we consider two probability distributions $\mathcal{P}=\{p_i\}^{n}_{i=1}$ and $\mathcal{Q}=\{q_i\}^{n}_{i=1}$. The Tsallis divergence between them can be defined as \cite{Nielsen2011OnRA}

\begin{align}\label{tsallisdiv}
	D_{\beta}(\mathcal{P}||\mathcal{Q})=\sum_{i=1}^{n} \frac{p_i^{\beta}q_i^{1-\beta}-1}{\beta-1}, 
\end{align}
where $\beta\in \mathbb{R}$ and limit has to be taken for $\beta$ tending to 1.
\subsubsection{Measures from $U^{\uparrow}(\mathcal{P})$}
On substituting the Tsallis divergence in Eq.(\ref{disuncer}), we get the Tsallis entropy as the measure of uncertainty\cite{tsallis1988possible}.

\begin{align}\label{tsallisent}
U^{\uparrow}_{\beta}(\mathcal{P})&=D_{\beta}(\mathcal{P}_C||\mathcal{P}_U)-D_{\beta}(\mathcal{P}||\mathcal{P}_U) \nonumber \\
&=\frac{1}{(\beta-1)n^{1-\beta}}\Big(1-\sum_{i=1}^n p_i^{\beta}\Big). 
\end{align}

\subsubsection{Measures from $U^{\downarrow}(\mathcal{P})$} 
If we substitute the Tsallis divergence in Eq.(\ref{disuncer2}), we get 

\begin{align*}
U^{\downarrow}_{\beta}(\mathcal{P})&=D_{\beta}(\mathcal{P}_U||\mathcal{P}_C)-D_{\beta}(\mathcal{P}_U||\mathcal{P}) \nonumber \\
&=\frac{1}{n^{\beta}(\beta-1)}(1-n^{\beta}-\sum_{i=1}^np_i^{1-\beta}).
\end{align*}
This measure is well defined only for $\beta\in(0,1)$. It can be easily observed that above measure can be obtained from Tsallis uncertainty by adding a constant term followed by rescaling.

\subsection{Hellinger distance}
Hellinger distance between $\mathcal{P}=\{p_i\}^{n}_{i=1}$ and $\mathcal{Q}=\{q_i\}^{n}_{i=1}$ is defined as \cite{le2012asymptotic} 
\begin{align*}
D_H(\mathcal{P}||\mathcal{Q})=\sum_{i}(||\sqrt{p_i}-\sqrt{q_i}||)^2.
\end{align*}
As this is a symmetric divergence measure, it will give same measure of uncertainty via both Eqs. (\ref{disuncer} and \ref{disuncer2}). By using Hellinger distance in Eq.(\ref{disuncer}), we get the following measure of uncertainty
\begin{align}\label{uhellinger}
U_H(\mathcal{P}) &= D_H(\mathcal{P}_C||\mathcal{P}_U)- D_H(\mathcal{P}||\mathcal{P}_U) \nonumber \\ 
&=\frac{1}{n}+1-\frac{2}{\sqrt{n}}-\sum_ip_i^2-\frac{1}{n}+\frac{2}{\sqrt{n}} \nonumber \\
&=1-\sum_{i}p_i^2.
\end{align}
We see that this reproduces the rescaled Tsallis entropic measure of uncertainty with $\beta=2$ and $n=1$ in Eq.(\ref{tsallisent}).
\subsection{The Total variation distance}
The Total variation distance between two probability distributions $\mathcal{P}$ and $\mathcal{Q}$ on $\mathcal{E}$ is defined as 
\begin{align*}
D_{TV}(\mathcal{P}||\mathcal{Q})=\supp_{A\subset \mathcal{E}}|\mathcal{P}(A)-\mathcal{Q}(A)|.
\end{align*}
Intuitively, it is the largest possible difference between two distributions on $\mathcal{P}(\mathcal{E})$, set of probabilities on $\mathcal{E}$.

For a finite or countable $\mathcal{E}$, this reduces to $\frac{1}{2}$ times the $l_1$-norm \cite{Aldous2019}
\begin{align*}
D_{TV}(\mathcal{P}||\mathcal{Q})=\frac{1}{2}\sum_i|p(i)-q(i)|,
\end{align*}
which is symmetric with respect to the probabilities. Hence, this will give same measure of uncertainty from  Eqns.(\ref{disuncer}) and (\ref{disuncer2}).
Using the total variation as the measure of statistical divergence in Eqn.(\ref{disuncer}), we get the following 
\begin{align}\label{utv}
U_{TV}(\mathcal{P})&=D_{TV}(\mathcal{P}_C||\mathcal{P}_U)-D_{TV}(\mathcal{P}||\mathcal{P}_U) \nonumber \\
&=1-\frac{1}{n}- \frac{1}{2}\sum_{i}|\frac{1}{n}-p_i|.
\end{align}
The above quantity is a new measure of uncertainty, which we name as ``Absolute uncertainty". We will also discuss its properties in the next Section. 

\begin{figure}[htb]
	\centering
	\includegraphics[scale=0.9]{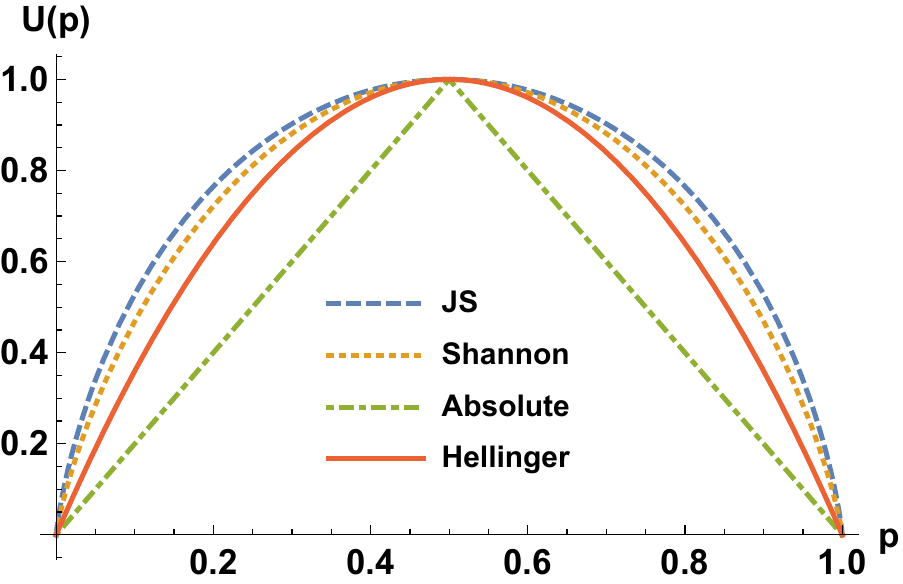}
	\caption{(Color online) Numerical plots of Shannon, Jensen-Shannon (JS), Absolute, and Hellinger measures of uncertainty. The Absolute measure of uncertainty has a discountinuity in its derivative at the maximally uncertain point. For fair comparison, we normalize the Hellinger and Absolute uncertainty.}
	\label{fig:b}
\end{figure}

In Fig.(\ref{fig:b}), we plot Shannon, Jensen-Shannon, Absolute, and Hellinger uncertainty measures for a two dimensional probability distribution $\{p,1-p\}$ with the parameter $p$. This figure shows that all the measures are faithful and continuous. However, Absolute uncertainty shows discontinuity in its first derivative at the maximum uncertain point.

\section{Properties of new Uncertainty measures}\label{sec4}
We will discuss the properties of the Jensen-Shannon uncertainty $U_{JS}(\mathcal{P})$ and absolute uncertainty $U_{TV}(\mathcal{P})$. A general discussion on the properties of various entropic uncertainty measures can be found in \cite{e10030261,ILIC2014138,ILIC2014138,1317122}.

\subsection{Jensen-Shannon uncertainty}

\begin{itemize}
\item	\textbf{Continuity}-It can be easily seen that the quantity $U_{JS}(\mathcal{P})$ in Eq.({\ref{jenshan}}) is a continuous function of the allowed values of $p_i$'s.
\item \textbf{Maximality}- As $U_{JS}(\mathcal{P})$ is a Schur-concave non-negative function, it attains its maximum value for the uniform distribution $\mathcal{P}_U$.
\item  \textbf{Expandibility}- Expandability means that $U(p_1,...,p_n)=U(p_1,...,p_n,0)$. Unlike Renyi and Tsallis entropies, the Jensen-Shannon is not expandable as $U_{JS}(\mathcal{P})$ has dimension dependent terms. One can check this for the simplest case of $\mathcal{P}=\{p_1,p_2\}$ as follows 
\begin{align*}
	U_{JS}(p_1,p_2)=\log\Big(\frac{16}{\sqrt{27}}\Big)+S(\mathcal{P})-S\Big(\frac{2\mathcal{P}+1}{2}\Big)\\ 
	U_{JS}(p_1,p_2,0)=\log\Big(\frac{36}{4^{\frac{4}{3}}}\Big)+S(\mathcal{P})-S(\frac{3\mathcal{P}'+1}{2}),
\end{align*}
where $\mathcal{P}'=\{p_1,p_2,0\}$.

 We argue here that expandability property need not be a necessary condition for an uncertainty measure. For example, consider a two dimensional probability distribution $\mathcal{P}_2=\{\frac{1}{2},\frac{1}{2}\}$ which is expanded to a three dimensional probability distribution $\mathcal{P}_3 =\{\frac{1}{2},\frac{1}{2},0\}$. While $\mathcal{P}_2$ is itself the uniform distribution in two dimensions, $\mathcal{P}_3$ is far away from the uniform distribution $\{\frac{1}{3},\frac{1}{3},\frac{1}{3}\}$ in three dimensions. Thus, it is not natural to expect expandability in this scenario.

\item \textbf{Additivity}- Additivity means that for the probability distributions, $\mathcal{P}=\{p_1,p_2,...,p_n\}$, $\mathcal{R}=\{r_{11},r_{12},...,r_{nm}\}$ $\forall n,m\in \mathbb{N}$ if there exists a probability distribution $\mathcal{Q}=\{q_{1|k},q_{2|k},...,q_{m|k}\}$, where $r_{ij}=q_{i|k}p_i$, then we can express the uncertainty in $\mathcal{R}$ as following

\begin{align*}
	U(\mathcal{R})=U(\mathcal{P})+U(\mathcal{Q}|\mathcal{P}).
\end{align*}

Instead, for the Jensen-Shannon uncertainty, we have
\begin{align*}
	U_{JS}(\mathcal{R})&=\log\Big(\frac{4(nm)^2}{(n+1)^{\frac{1+nm}{nm}}}\Big)-\sum_{i=1}^{nm}r_i\log(r_i)\nonumber \\ &+\frac{2}{nm}\sum_{i=1}^{nm}\frac{nmr_i+1}{2}\log\Big(\frac{nmr_i+1}{2}\Big).
\end{align*}

Clearly, except for the second term, no other terms can be written as linear addition.
\end{itemize}

\subsection{Absolute uncertainty}

\begin{itemize}
	\item \textbf{Continuity}-Again, it is easy to check that the quantity $U_{TV}(\mathcal{P})$ in Eq.({\ref{utv}}) is a continuous function of the allowed values of $p_i$'s.
	\item \textbf{Maximality}- As $U_{TV}(\mathcal{P})$ is a Schur-concave non-negative function, it attains its maximum value for the uniform distribution $\mathcal{P}_U$.
	\item \textbf{Expandibility}- Similar to Jensen-Shannon uncertainty, the Absolute uncertainty is not expandable as $U_{TV}(\mathcal{P})$ has dimension dependent terms. One can again find it via the simple example of  $U_{TV}(p_1,p_2)$ to $U_{TV}(p_1,p_2,0)$ as following 
	\begin{align*}
		&U_{TV}(p_1,p_2)=\frac{1}{2}-\frac{1}{2}\Big(\big|\frac{1}{2}-p_1\big|+\big|\frac{1}{2}-p_2\big|\Big),  \\
		&U_{TV}(p_1,p_2,0)=\frac{5}{9}-\frac{1}{3}\Big(\big|\frac{1}{3}-p_1\big|+\big|\frac{1}{3}-p_2\big|\Big).
	\end{align*}
	\item \textbf{Additivity}- For the Absolute uncertainty we have 
	\begin{align*}
		U_{TV}(\mathcal{R})=1-\frac{1}{nm}- \frac{1}{2}\sum_{i}|\frac{1}{nm}-r_i|
	\end{align*}

Here also, we can not expand it as a linear sum of $U_{TV}(\mathcal{P})$ and $U_{TV}(\mathcal{Q}|\mathcal{P})$ as the terms inside the modulus can not be separated.
\end{itemize}

%

\section{Uncertainty in the preparation of a quantum state}\label{sec5}
Here, we will discuss the analogical extension of the above formalism to quantum domain. Let us consider that a finite dimensional quantum system is described by density matrix $\rho\in {\mathcal L}({\mathcal H})$. If $\rho^2\neq \rho$, it is a mixed state or in other words we say that it has preparation uncertainty. For pure state, the uncertainty is zero while it is maximum for maximally mixed state. Now, question is: how far our $\rho$ is from the pure states, will be its preparation uncertainty. But finding this distance requires a minimization over all pure states. To bypass this, one can consider the following distances, $\mathcal{D} \left(V\ketbra{\psi}{\psi}V^{\dagger}|| \frac{\mathbb{I}}{d}\right)$ and $\mathcal{D} \left(\rho|| \frac{\mathbb{I}}{d}\right)$, then finally can reach to 
\begin{align}
U(\rho)=\mathcal{D} \left(V\ketbra{\psi}{\psi}V^{\dagger}|| \frac{\mathbb{I}}{d}\right)-\mathcal{D} \left(\rho|| \frac{\mathbb{I}}{d}\right),
\end{align}
where $\mathcal{D} (\rho||\sigma)$ is a valid distance measure between two density matrices, $V$ is an arbitrary $d\times d$ unitary matrix and $d$ is dimension of the Hilbert space of the states. Now we will consider some known statistical divergences in the quantum case and see what measures of uncertainty they will induce. 

\subsection{Bures distance and Hellinger distance}
Bures and Hellinger distance for two density matrices $\rho, \sigma \in {\mathcal L}({\mathcal H})$ are defined as \cite{chuang,PhysRevA.69.032106}
\begin{align*}
D_B(\rho ||\sigma)=2-2F(\rho,\sigma) \:\:\mbox{and}\:\: D_H(\rho ||\sigma)=2-2A(\rho,\sigma)
\end{align*}
respectively, where $F(\rho,\sigma)={\rm Tr}\sqrt{\sqrt{\rho}\sigma\sqrt{\rho}}$ is Fidelity and $A(\rho,\sigma)={\rm Tr}\sqrt{\rho}\sqrt{\sigma}$ is Affinity between two states $\rho,\sigma$. It can be seen that the both distances induce the same uncertainty measure, i.e., 
\begin{align}
U(\rho)=\frac{1}{\sqrt{d}}({\rm Tr}\sqrt{\rho}-1),
\label{B-Hell}
\end{align}
by noticing that $F(\mathbb{I}/d,V\ketbra{\psi}{\psi}V^{\dagger})=A(\mathbb{I}/d,V\ketbra{\psi}{\psi}V^{\dagger})=\frac{1}{\sqrt{d}}$ and $F(\mathbb{I}/d,\rho)=A(\mathbb{I}/d,\rho)=\frac{1}{\sqrt{d}}{\rm Tr}\sqrt{\rho}$. The uncertainty measure in Eq.(\ref{B-Hell}) is similar to linear entropy and related to Tsallis entropy $T_{\frac{1}{2}}(\rho)$. 

\subsection{Distance induced by $l_p$-norm and Schatten $p$-norm}
For a $d_1\times d_2$ matrix $M=\{M_{ij}\}$ and $p\in [1,\infty)$, the two norms are defined as 
\begin{align*}
||M||_{l_p}=\left(\sum_{i,j}|M_{ij}|^p\right)^{\frac{1}{p}}\:\:\mbox{and}\:\: ||M||_{p}=\left(\sum_{i}^r \lambda_i^p\right)^{\frac{1}{p}},
\end{align*}
where $\lambda_i$ are non-zero eigen values of $|M|=\sqrt{M^{\dagger} M}$ and $r$ is the rank of $M$. Now the distance induced by these two norms are, respectively, $D_{l_p}(\rho ||\sigma)=||\rho-\sigma||_{l_p}$ and $D_{p}(\rho ||\sigma)=||\rho-\sigma||_{p}$. We find that these two distances yield same information measure as
\begin{align}
U(\rho)=\frac{[(d-1)^p+d-1]^{\frac{1}{p}}}{d}-\left(\sum_{i=1}^d|\lambda_i^{\rho}-\frac{1}{d}|^p\right)^{\frac{1}{p}},
\label{lp-p}\end{align}
where $\lambda_i^{\rho}$ are the eigenvalues of $\rho$.

\subsection{Hilbert-Schmidt distance}
Hilbert-Schmidt distance is induced by Hilbert-Schmidt norm, and for two density matrices $\rho, \sigma$, it is defined as 
\begin{align*}
\mathcal{D} _{HS}(\rho ||\sigma)={\rm Tr} [(\rho-\sigma)^2].
\end{align*}
We notice that $\mathcal{D} _{HS}(V\ketbra{\psi}{\psi}V^{\dagger}|| \mathbb{I}/d)=1-1/d$ and $\mathcal{D}_{HS} (\rho ||\mathbb{I}/d)={\rm Tr}\rho^2-1/d$. This tells us that the induced uncertainty measure is given by
\begin{align}
U_{HS}(\rho)=1-{\rm Tr}\rho^2,
\end{align}
which we all recognise as linear entropy.

\subsection{Generalized R\'{e}nyi divergence}
The generalized R\'{e}nyi divergence was introduced in \cite{muller2013quantum} and is defined as 
\begin{align*}
D_{\alpha}(\rho ||\sigma)=\frac{1}{\alpha-1}\log{\rm Tr}\left(\sigma^{\frac{1-\alpha}{2\alpha}}\rho \sigma^{\frac{1-\alpha}{2\alpha}}\right)^{\alpha},
\end{align*}
where $\alpha\in\mathbb{R}$ and $\rho$ is not orthogonal to $\sigma$. 
Whenever, $\rho$ and $\sigma$ commute,  the generalized R\'{e}nyi divergence reduces to classical $\alpha$-R\'{e}nyi divergence.
Clearly, for our case, as maximally mixed state commutes with both $\rho$ and $V\ketbra{\psi}{\psi}V^{\dagger}$, the distances, $D_{\alpha}(V\ketbra{\psi}{\psi}V^{\dagger}||\mathbb{I}/d)=\log d$ and $D_{\alpha}(\rho ||\mathbb{I}/d)=\log d-\frac{1}{\alpha-1}\log \Tr \rho^{\alpha}$. Hence, we reach
\begin{align}
 U_{\alpha}(\rho)=\frac{1}{1-\alpha}\log \Tr \rho^{\alpha}=S_{\alpha}(\rho), 
\end{align}
 the well known R\'{e}nyi entropy.
 
 \begin{figure}[htb]
 	\centering
 	\includegraphics[scale=0.9]{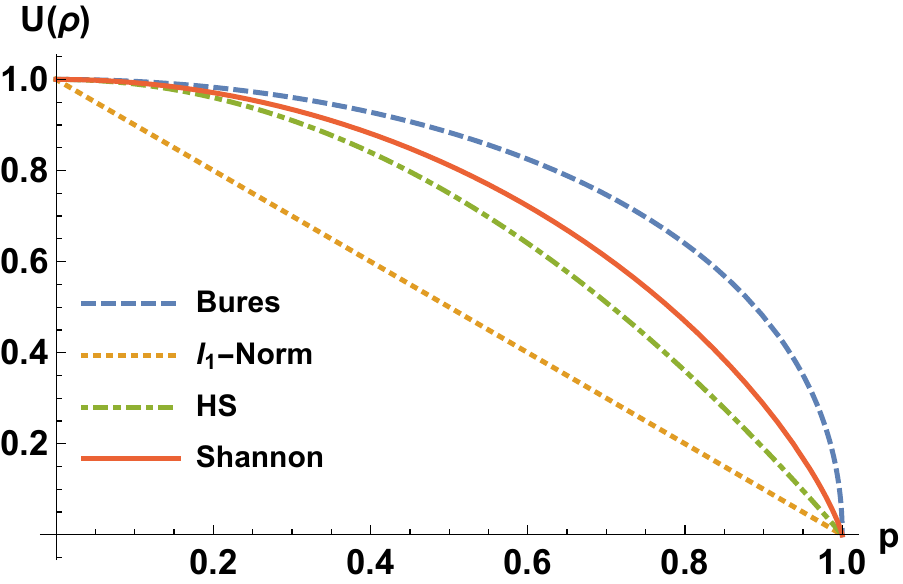}
 	\caption{Numerical plot of Bures (Hellinger), $l_1$-norm, Hilbert-Schmidt (HS), and Shannon measures of uncertainty for the state, $\rho=p\ketbra{\psi}{\psi}+(1-p)\frac{\mathbb{I}}{2}$, where $\ket{\psi}$ is an arbitrary pure state in $d=2$. For fair comparison, we normalize the uncertainty induced by Bures and Hellinger distances.}
 	\label{fig:c}
 \end{figure}

\subsection{Generalized Tsallies divergence}
The Tsallis divergence is defined as \cite{PhysRevA.89.012331}
\begin{align}
D_{\beta}(\rho ||\sigma)=\frac{1}{1-\beta}{\rm Tr}\left(\sigma^{\frac{1-\beta}{2}}\rho^{\frac{\beta}{}} \sigma^{\frac{1-\beta}{2}}\right). 
\end{align}
Then, $D_{\beta}(V\ketbra{\psi}{\psi}V^{\dagger}||\mathbb{I}/d)=\frac{1}{1-\beta}(\frac{1}{d^{1-\beta}}-1)$ and $D_{\beta}(\rho||\mathbb{I}/d)=\frac{1}{1-\beta}(\frac{1}{d^{1-\beta}}{\rm Tr}[\rho^\beta]-1)$. Thus, 
\begin{align}
U_{\beta}(\rho)= \frac{-1}{d^{1-\beta}} \left[\frac{1}{1-\beta}({\rm Tr}[\rho^\beta]-1)\right]=\frac{(-1)}{d^{1-\beta}}T_{\beta}(\rho), 
\end{align}
is the Tsallies entropy, $T_\beta(\rho)$ with a factor $\frac{-1}{d^{1-\beta}}$. 

We plot Bures (Hellinger), $l_1$-norm, Hilbert-Schmidt, and Shannon uncertainty measures in Fig.(\ref{fig:c}) for an arbitrary state $\rho$ in $d=2$. 
The figure hints that the uncertainty captured by $l_1$-norm measure is lowest for $0< p< 1$ and Bures measure upper bound the others.

\section{Conclusion} \label{sec6}
In this work we have shown how the uncertainty measures arise from the geometry of statistical divergence measures. It captures the essence of uncertainty as the distance of a probability distribution form a certain probability distribution. We also report two new forms of uncertainty from the Jensen-Shannon divergence and Total Variation distance and discuss their properties. In particular, the two new uncertainty measures do not satisfy the expandability axiom, which is satisfied by the more commonly used uncertainty measures.

We also apply a similar geometric technique to obtain the uncertainty in the preparation of a state or the mixedness. We reproduce the commonly used measures of mixedness using various distance measures of two quantum states. 

This work opens up several new directions of research. First, it would be interesting to see which other geometric approaches can produce uncertainty measures from divergences. Our work also sets up a standard method for finding various uncertainty measures. Second, it would also be important to find the uncertainty relations and various applications of the new uncertainty measures found here.

\textit{Acknowledgement}:-- SS acknowledges the financial support through the {\v S}tefan Schwarz stipend from Slovak Academy of Sciences, 
Bratislava. SS also acknowledges the financial support through the project OPTIQUTE (APVV-18-0518) and HOQIP (VEGA 2/0161/19). 
\bibliography{uncertaintyref}
\end{document}